\documentclass[11pt,a4paper]{article}

\usepackage[utf8]{inputenc}
\usepackage{amsmath}
\usepackage{amsthm}
\usepackage{amssymb}
\theoremstyle{plain}
\newtheorem{theorem}{Theorem}
\newtheorem{lemma}{Lemma}

\usepackage{graphicx}
\usepackage{bbold}
\usepackage{fancybox}
\usepackage{bm}
\usepackage{float}
\usepackage{geometry}
\usepackage{multicol}
\usepackage{multirow}
\usepackage{booktabs}
\usepackage{bbding}
\usepackage{comment}
\usepackage{verbatim}
\usepackage[round]{natbib}
\usepackage{setspace}
\usepackage{caption}
\usepackage{threeparttable}

\restylefloat{table}

\makeatletter
\renewenvironment{abstract}
  {\begin{center}
   \bfseries\normalsize \abstractname
   \end{center}
   \quotation\normalsize}
  {\endquotation}
\makeatother

\begin{document}

\title{Identification in Auctions with Truncated Transaction Prices\footnote{I am grateful to Professor Sorawoot Srisuma for his guidance and detailed comments. I also thank Professor Gregory Cox, Ming Li, Jingfeng Lu, Satoru Takahashi, and Allen Vong for their valuable suggestions. All remaining errors are my own.} }
\author{Tonghui Qi\footnote{Address: Room 413B, Mingde Main Building, 59 Zhongguancun Street, Haidian District, Beijing 100872, China; E-mail: qitonghui@ruc.edu.cn.} \\ Renmin University of China}
\date{\today}

\maketitle

\begin{abstract}
This paper studies nonparametric identification in auction models with truncated transaction price data arising from either binding reserve prices or endogenous entry. I characterize identification under alternative information structures that differ according to whether the number of potential bidders is fixed or varies across auctions and whether the econometrician observes the number of potential or active bidders. A central insight is that identification hinges on recovering the screening threshold that determines bidder participation. The analysis provides a unified characterization of identification in first- and second-price auctions under both truncation mechanisms. \\


\end{abstract}

\maketitle

\newpage

\section{Introduction}

Identification in auctions has been studied extensively over the past three decades, with the bidder value distribution being one of the central objects of interest. Within the independent private values (IPV) paradigm, the seminal papers by \cite{guerre2000optimal} and \cite{athey2002identification} establish nonparametric identification of the bidder value distribution from bids in first-price and second-price auctions, respectively.

Building on these seminal contributions, a large literature has extended identification in auction models well beyond the symmetric IPV framework. Examples include affiliated or correlated private values \citep{li2002structural, aradillas2013identification}, asymmetric bidders \citep{campo2003asymmetry,somaini2020identification,
bhattacharya2022semiparametric}, risk-averse bidders \citep{guerre2009nonparametric}, unobserved auction heterogeneity \citep{krasnokutskaya2011identification, hu2013identification, luo2023identification}, and interdependent values \citep{fevrier2008nonparametric,gimenes2020nonparametric}. Despite these advances, the existing literature almost universally assumes that bids or transaction prices are observed in every auction.

This assumption, however, may fail in many empirical settings. When a seller sets a binding reserve price, bidders whose values lie below the reserve do not submit bids, and auctions in which all bidders are screened out generate neither bids nor transaction prices. A similar problem arises when participation is costly. If entering an auction requires paying a fixed cost, only bidders with sufficiently high values choose to participate, and auctions in which no bidder enters likewise generate no observable data. In both cases, the econometrician observes transaction prices only when at least one bidder submits a bid, giving rise to truncated transaction price data.

Truncation creates two distinct identification challenges. First, the observed number of bidders corresponds only to those who submit bids (active bidders), rather than the total number of potential bidders who could have participated. Second, transaction prices are observed only in auctions with at least one active bidder, while auctions in which all potential bidders are screened out remain unobserved. Consequently, the observed transaction prices are drawn from a truncated distribution, and ignoring this truncation generally leads to biased and inconsistent estimation of the bidder value distribution.

To avoid the truncation problem, some empirical studies assume that reserve prices are nonbinding; see \cite{campo2011semiparametric}, \cite{haile2003inference}, and \cite{aradillas2013identification} in the context of U.S. Forest Service timber auctions. In many other auction environments, however, reserve prices are binding \citep{roberts2013unobserved, choi2016reserve, ostrovsky2023reserve}, making truncation an inherent feature of the observed data. Traditional identification methods typically rely on the unconditional distribution of bids or transaction prices together with information on the number of potential bidders, and therefore may no longer apply. Despite the empirical relevance of this problem, relatively little is known about identification from truncated transaction price data. A notable exception is \cite{shneyerov2011identification}, who study first-price auctions with binding reserve prices under the assumption that the number of active bidders is observed. This paper complements their analysis by providing a unified characterization of identification across a broader class of auction formats and information structures.

Throughout this paper, I assume that only transaction prices—rather than the full set of bids—are observed, and that either the number of potential bidders or the number of active bidders is observed, but not both. These assumptions are empirically relevant. While observing all bids necessarily reveals the transaction price, the converse is not true, and in many auction settings only the final transaction price is recorded. Moreover, in online auctions, only active bidders are typically observed, whereas in other settings institutional features may reveal the number of potential bidders even though only a subset ultimately participates. Under this limited information structure, I characterize when the bidder value distribution is point identified, set identified, or not identified. Since the transaction price distribution in descending and ascending auctions coincides with that in first- and second-price auctions, respectively \citep{milgrom1982theory}, I focus on first- and second-price auctions, although the identification results apply directly to descending and ascending auctions.

To characterize identification under truncated transaction price data, I consider a collection of information structures that differ along two dimensions. The first is whether the number of potential bidders is fixed across auctions or varies exogenously. The second is whether the econometrician observes the number of potential bidders or only the number of active bidders. Together, these information structures capture the data environments most commonly encountered in empirical auction studies and determine the information available to recover the bidder value distribution.

I first study auctions with binding reserve prices. When the number of potential bidders is fixed across auctions, second-price auctions are point identified, whereas first-price auctions additionally require information on the frequency of no-sale auctions. I then allow the number of potential bidders to vary across auctions. If the number of potential bidders is observed, its exogenous variation is sufficient to achieve point identification in both first- and second-price auctions. If the number of potential bidders varies across auctions but is unobserved, while the number of active bidders is observed, the identification pattern is reversed. First-price auctions are point identified, whereas second-price auctions are generally not identified. In the latter case, observing the frequency of no-sale auctions yields a finite identified set.

I then extend the analysis to auctions with entry costs, where truncation arises from endogenous participation rather than a binding reserve price. Despite the different economic mechanisms, the two environments exhibit remarkably similar identification structures. In both cases, identification depends on recovering the screening threshold that governs bidder participation, although the threshold is determined differently. Under binding reserve prices, the screening threshold is invariant across auctions under the maintained assumptions, whereas under entry costs it varies endogenously with the number of potential bidders. This distinction gives rise to different identifying arguments while preserving a common identification framework. Table \ref{tab:identification_summary} summarizes the identification results under the alternative information structures considered in the paper, where $T^{\text{obs}}$ denotes the observed transaction price, $N$ denotes the number of potential bidders, $N^+$ denotes the observed number of active bidders, and $L^{\text{invalid}}$ denotes the number of no-sale auctions. In the table, “Point ID” denotes point identification, “Finite Set ID” indicates that the identified set contains finitely many elements, and “Not ID” indicates that the model is not identified, in the sense that infinitely many parameter values are observationally equivalent.\footnote{The “Point ID” result for second-price auctions in the last row assumes $\operatorname{supp}(N)=\{2,3,\ldots,\bar N\}$, as in Theorem 10.}

\begin{table}[ht] 
\centering 
\caption{Identification under Alternative Information Structures} \label{tab:identification_summary} 
\begin{tabular}{lllcc} 
\toprule 
Source of truncation & $N$ type & Observed data & First-price & Second-price \\ 
\midrule 
\multirow{5}{*}{Binding reserve} & Fixed & $T^{\mathrm{obs}},\,N$ & Not ID & Point ID \\ 
& Fixed & $T^{\mathrm{obs}},\,N,\,L^{\mathrm{invalid}}$ & Point ID & Point ID \\ 
& Varying & $T^{\mathrm{obs}},\,N$ & Point ID & Point ID \\ 
& Varying & $T^{\mathrm{obs}},\,N^{+}$ & Point ID & Not ID \\ 
& Varying & $T^{\mathrm{obs}},\,N^{+},\,L^{\mathrm{invalid}}$ & Point ID & Finite Set ID \\ 
\midrule 
\multirow{5}{*}{Entry cost} & Fixed & $T^{\mathrm{obs}},\,N$ & Not ID & Point ID \\ 
& Fixed & $T^{\mathrm{obs}},\,N,\,c$ & Point ID & Point ID \\ 
& Varying & $T^{\mathrm{obs}},\,N$ & Finite Set ID & Point ID \\ 
& Varying & $T^{\mathrm{obs}},\,N^{+}$ & Point ID & Not ID \\ 
& Varying & $T^{\mathrm{obs}},\,N^{+},\,c$ & Point ID & Point ID \\ 
\bottomrule 
\end{tabular} 
\end{table} 

The remainder of the paper is organized as follows. Section 2 reviews the related literature. Section 3 introduces the model and the information structures. Section 4 presents the identification results under binding reserve prices. Section 5 extends the analysis to auctions with entry costs. Section 6 concludes.

\section{Related Literature}

This paper relates to three strands of the auction literature. The first studies identification under binding reserve prices. Although binding reserve prices have been considered in several identification studies, the implications of truncated transaction price data have not been systematically characterized across different auction formats and information structures. This paper provides such a characterization.

The existing literature establishes identification under a variety of information structures. \cite{guerre2000optimal} show that in first-price auctions with a fixed and known number of potential bidders, the bidder value distribution is identified when the number of active bidders and the bid distribution above the reserve price are observed. I complement their analysis by considering alternative information structures, particularly environments in which the number of potential bidders varies across auctions. In second-price auctions, \cite{athey2002identification} show that when the number of potential bidders is fixed and known, the transaction price identifies the bidder value distribution conditional on values exceeding the reserve price. I further characterize when the unconditional bidder value distribution is identified under alternative information structures. More recently, \cite{shneyerov2011identification} establish identification in first-price auctions with binding reserve prices using bid data together with the number of active bidders. In contrast, I characterize identification in both first- and second-price auctions while allowing either the number of potential bidders or the number of active bidders to be observed.

The second strand studies auctions with an unknown number of bidders. Existing work primarily investigates how the bidder value distribution can be recovered when the number of potential bidders is unobserved but richer bidding information is available. For example, \cite{song2004nonparametric} shows that in ascending auctions with IPV bidders, observing any two order statistics of valuations is sufficient for nonparametric identification. \cite{an2010estimating} develop a nonparametric procedure to recover the bid distribution conditional on the unknown number of potential bidders in first-price auctions. \cite{freyberger2022identification} study ascending auctions with unobserved heterogeneity and an unknown number of bidders, deriving different identification results depending on whether the reserve price is public or secret. Unlike these papers, I assume that only the transaction price is observed. \cite{adams2007estimating} also relies solely on transaction prices and shows that the bidder value distribution is identified when the distribution of the number of potential bidders is known. By contrast, I allow this distribution to be unknown and characterize the additional information required for identification.

The third strand studies identification and estimation in auctions with entry costs. Existing work has considered both identification and estimation under a variety of entry models. \cite{li2005econometrics} develops a simulated moments estimator for first-price auctions with endogenous entry, whereas this paper studies nonparametric identification. Another branch of the literature considers two-stage entry models in which bidders first observe private signals, decide whether to incur an entry cost, and learn their valuations only after entering \citep{marmer2013model, gentry2014identification, chen2026identification}. In contrast, I adopt a Samuelson-type environment \citep{samuelson1985competitive}, in which bidders observe their valuations before making entry decisions. The paper most closely related to my analysis is \cite{xu2013nonparametric}, which studies nonparametric identification in a Samuelson-type auction model with entry costs. The key difference is that \cite{xu2013nonparametric} assumes both the number of potential bidders and the number of active bidders are observed, whereas I consider the empirically relevant setting in which only one of these two margins is available.

Taken together, the existing literature studies binding reserve prices, unknown numbers of bidders, and entry costs largely in isolation. This paper unifies these strands by characterizing identification under truncated transaction price data across a common set of information structures.

\section{Model and Information Structures}

This section introduces the notation and information structures used throughout the paper. I begin with a simple example that illustrates the identification problem before presenting the formal framework.

\subsection{Motivating Example}

The identification problem can be illustrated through a simple example. Consider a second-price auction with a public reserve price. Since truthful bidding is a weakly dominant strategy in second-price auctions, bids coincide with bidders' private values, allowing the truncation mechanism to be illustrated without introducing equilibrium bidding strategies. Consider three auctions, each with two potential bidders, whose private values are given in Table \ref{tab:motivating_example}. The seller sets a public reserve price of 2.5. The table compares the information observed by the econometrician with and without the reserve price, where $T$ denotes the transaction price and $N^{\mathrm{act}}$ denotes the number of active bidders.

\begin{table}[H]
\centering
\caption{A motivating example}
\label{tab:motivating_example}
\begin{tabular}{lccc}
\toprule
Auction &
Private values &
No reserve &
Reserve price = 2.5 \\
\midrule
1 & $(3,4)$ & $T=3,\;N^{\mathrm{act}}=2$ & $T=3,\;N^{\mathrm{act}}=2$ \\
2 & $(2,3)$ & $T=2,\;N^{\mathrm{act}}=2$ & $T=2.5,\;N^{\mathrm{act}}=1$ \\
3 & $(1,2)$ & $T=1,\;N^{\mathrm{act}}=2$ & $T$ unobserved, $N^{\mathrm{act}}=0$ \\
\bottomrule
\end{tabular}
\end{table}

Several observations are worth noting. Auction 1 is unaffected by the reserve price because both bidders' values exceed the reserve. In Auction 2, only one bidder's value exceeds the reserve. The auction remains valid, but the transaction price increases to the reserve price and only one bidder remains active. In Auction 3, both bidders are screened out by the reserve price, and the auction ends without a sale.

This simple example illustrates the central identification problem studied throughout the paper. Relative to the standard setting without truncation, the econometrician loses information along two dimensions. First, transaction prices are observed only for auctions that result in a sale. Second, the observed number of bidders corresponds only to active bidders rather than potential bidders. The remainder of the paper asks under what information structures the bidder value distribution can nevertheless be recovered from such truncated transaction price data.

\subsection{Observables and Information Structures}

The identification results in this paper depend critically on the information available to the econometrician. Throughout the paper, I assume that only transaction prices, rather than the full vector of bids, are observed. This assumption is empirically relevant: while observing all bids necessarily reveals the transaction price, the converse is not true, and many auction datasets record only the final transaction price.

Let $T^{\mathrm{obs}}$ denote the transaction price observed by the econometrician. In the presence of a binding reserve price $R$, it is defined as
\[
T^{\mathrm{obs}}=
\begin{cases}
b^{N:N}\mid b^{N:N}\geq R, 
& \text{in first-price auctions},\\[6pt]
\max\{b^{N-1:N},R\}\mid b^{N:N}\geq R,
& \text{in second-price auctions}.
\end{cases}
\]
That is, the transaction price is observed only when at least one bidder's bid exceeds the reserve price. When all bidders are screened out by the reserve price, the auction generates no transaction price and is referred to as an invalid auction.

In addition to transaction prices, information about bidder participation may also be available. Let $N$ denote the number of potential bidders, namely, the total number of bidders who are eligible to participate in the auction. Let $N^{\mathrm{act}}$ denote the number of active bidders, namely, the number of bidders who actually participate by submitting bids. Since the number of active bidders is observed by the econometrician only for valid auctions, the number of observed active bidders is denoted by
\[
N^+=N^{\mathrm{act}}\mid N^{\mathrm{act}}\geq 1.
\]
For invalid auctions, $N^{\mathrm{act}}=0$, so neither the transaction price nor the number of active bidders is observed.

The distinction between potential and active bidders is central to the identification problem studied in this paper. In the absence of truncation, these two measures of bidder participation coincide. However, binding reserve prices and entry costs may prevent some potential bidders from participating, causing the econometrician to observe only active bidders rather than the underlying number of potential bidders.\footnote{See \cite{hendricks2007empirical} for a discussion of the empirical relevance of this assumption.}

The paper studies identification under alternative information structures characterized by two dimensions. The first dimension concerns whether the number of potential bidders is fixed across auctions or varies exogenously. The second dimension concerns which measure of bidder participation is observed. The econometrician may observe the number of potential bidders or only the number of active bidders, but not both. These two dimensions determine the amount of information available for recovering the bidder value distribution from truncated transaction prices.

\subsection{Screening and Quantile Representation}

Both binding reserve prices and entry costs generate truncation through a screening mechanism. In a binding reserve price environment, bidders whose values are below the reserve price do not participate in the auction. Similarly, when participation requires an entry cost, bidders with sufficiently low values choose not to enter. In both cases, the econometrician observes transaction prices only after bidders have passed through this screening process.

The key object underlying the identification analysis is the screening threshold. Let $F$ denote the cumulative distribution function (CDF) of bidder values, and let $V(\alpha)=F^{-1}(\alpha)$ denote the corresponding quantile function. The screening threshold is represented by the quantile $\alpha^*$ such that bidders with values below $V(\alpha^*)$ are screened out. Equivalently, $\alpha^*$ represents the proportion of bidders whose values are below the participation threshold.

This quantile representation is useful because identification depends on recovering the quantile point at which screening occurs, rather than the screening threshold in levels. For example, under a binding reserve price, the reserve price corresponds to $V(\alpha^*)$, while the fraction of bidders screened out is summarized by $\alpha^*$. Expressing the model in terms of $\alpha^*$ therefore allows the identification analysis to focus on the fundamental truncation object and simplifies the notation throughout the paper.\footnote{The use of quantile-based representations in auction identification and estimation is well established; see, for example, \cite{gimenes2017econometrics} and \cite{gimenes2022quantile}.}

\section{Identification with Binding Reserve Prices}

This section studies identification when truncation arises from a binding reserve price. The key feature of this environment is that the reserve price creates a screening mechanism: bidders with values below the reserve price do not participate, and auctions in which all bidders are screened out generate no observed transaction price. The identification problem is therefore whether the available information is sufficient to recover the screening threshold and, consequently, the bidder value distribution above the reserve price. The following assumption characterizes the reserve-price environment considered in this section. \\

\textbf{Assumption R:}

\textit{(i) The support of $F(\cdot)$ is $[\underline{v},\bar{v}]$, where $0\leq \underline{v}<\bar{v}<\infty$.}

\textit{(ii) $F(\cdot)$ admits a continuous and strictly positive density $f(\cdot)$ on its support.}

\textit{(iii) The reserve price $R$ is constant across auctions and satisfies $R\in(\underline{v},\bar{v})$.}

\textit{(iv) All bidders play the symmetric and strictly increasing equilibrium strategy.} \\

Assumption R(i) restricts bidder values to be non-negative and bounded. Assumption R(ii) guarantees that $V'(\alpha)$ is well defined and strictly positive. Assumption R(iii) is the key assumption for identification in the presence of a binding reserve price. Since the reserve price is invariant across auctions, the corresponding screening threshold $\alpha^*=F(R)$ is also invariant. 
This invariance creates a common object linking auctions with different numbers of potential bidders and plays a crucial role in the identification analysis. Under standard regularity conditions, the optimal reserve price is independent of the number of potential bidders \citep{riley1981optimal}. Therefore, Assumption R(iii) is natural in environments where the seller is profit maximizing and the seller's outside option is stable across auctions. Finally, Assumption R(iv) imposes the standard symmetric equilibrium restriction used throughout the analysis.

Before presenting the identification results, it is useful to clarify the object of identification in the presence of a binding reserve price. First, the distribution of bidder values below the reserve price cannot be recovered from observed auction outcomes. Since bidders with values below $R$ do not participate, their value information is never revealed to the econometrician. Therefore, the identification analysis focuses on recovering $F(v)$ for $v\geq R$ rather than the entire value distribution.

Second, identifying $F(v)$ for $v\geq R$ is different from identifying the conditional distribution of bidder values given participation. Specifically,
\[
F(v\mid v\geq R)
=\frac{F(v)-F(R)}{1-F(R)}
=\frac{F(v)-\alpha^*}{1-\alpha^*},
\]
where $\alpha^*=F(R)$ denotes the screening threshold. Truncated transaction price data typically reveal the conditional distribution $F(v\mid v\geq R)$ rather than the unconditional distribution above the reserve price. Therefore, identification of $F(v)$ for $v\geq R$ requires additional information to recover $\alpha^*$. Equivalently, the identification problem can be formulated as recovering $(\alpha^*,V(\alpha))$ for $\alpha\geq\alpha^*$, which is the focus of the analysis below.

\subsection{Fixed Number of Potential Bidders}

I first consider the case where all auctions have the same number of potential bidders, which is known by the econometrician. This setting provides a benchmark where there is no uncertainty about the level of competition, and identification depends solely on whether the observed transaction prices contain sufficient information to recover the screening threshold. All proofs are given in the appendix. \\

\begin{theorem}
Suppose all auctions share a common number of potential bidders $N\geq 2$, which is known by the econometrician. Then $\alpha^*$ and $V(\alpha)$ for $\alpha\geq\alpha^*$ are point identified in second-price auctions but not in first-price auctions. \\
\end{theorem}

Theorem 1 highlights that the main challenge created by a binding reserve price is recovering the screening threshold $\alpha^*$. In second-price auctions, the observed transaction price distribution contains a mass point at the reserve price. This mass point arises because, whenever the highest bidder exceeds the reserve but the 
second-highest bidder does not, the winner pays the reserve price. The probability of this event identifies the fraction of bidders screened out by the reserve price and therefore identifies $\alpha^*$. Once $\alpha^*$ is recovered, the bidder value distribution above the reserve price can be obtained from the observed transaction price distribution.

In contrast, in first-price auctions, the observed transaction price distribution does not contain such a mass point, and thus does not provide sufficient information to recover $\alpha^*$. Intuitively, the observed transaction prices reveal only the distribution of bids among active bidders. Different screening thresholds and value distributions can generate the same distribution of observed transaction prices, leading to non-identification of the unconditional bidder value distribution above the reserve price.

The next theorem shows that observing additional information about unsuccessful auctions can restore identification in first-price auctions. \\

\begin{theorem}
Suppose all auctions share a common number of potential bidders $N\geq 2$, which is known by the econometrician. Then $\alpha^*$ and $V(\alpha)$ for $\alpha\geq\alpha^*$ are point identified in first-price auctions if the number of invalid auctions is observed. \\
\end{theorem}

Theorem 2 demonstrates the identifying value of no-sale auctions. The proportion of invalid auctions equals the probability that all potential bidders are screened out, which directly depends on the screening threshold $\alpha^*$. Therefore, observing the frequency of invalid auctions provides the additional information needed to recover 
$\alpha^*$. Once the screening threshold is identified, the bidder value distribution above the reserve price is also identified.

\subsection{Varying Number of Potential Bidders}

I next consider the case where the number of potential bidders varies across auctions. Unlike the fixed-$N$ case, variation in the number of potential bidders provides additional information because the screening threshold is invariant across auctions. This creates restrictions that link transaction price distributions across different levels of competition and can facilitate identification even when the transaction price distribution from a single value of $N$ is insufficient.

First, I consider the case where the number of potential bidders is observed for each auction and give the identification results in Theorem 3. \\

\begin{theorem}
Suppose the number of potential bidders $N$ is observed and takes at least two different values larger than one. Assume the bidder value distribution is invariant across auctions with different numbers of potential bidders. Then $\alpha^*$ and $V(\alpha)$ for $\alpha \geq \alpha^*$ are point identified in both first-price and second-price auctions. \\
\end{theorem}

Theorem 3 shows that exogenous variation in the number of potential bidders is sufficient to restore identification in both auction formats. The intuition is that the invariant screening threshold creates a common restriction across auctions with different levels of competition. In particular, the distribution of observed transaction prices changes with $N$, while the underlying truncation threshold remains unchanged. This cross-auction variation provides the additional information required to recover $\alpha^*$.

It is worth noting that identification in Theorem 3 does not require observing the mass point at the reserve price in second-price auctions. If the mass point is observed, the model becomes overidentified, and the assumption that the bidder value distribution is invariant across auctions with different numbers of potential bidders can be tested. 
Similarly, if $N$ takes more than two values, the additional variation provides further overidentifying restrictions.

In many empirical settings, however, the econometrician observes the number of active bidders rather than the total number of potential bidders. When the reserve price is binding, the number of active bidders may be strictly smaller than the number of potential bidders because some bidders are screened out. Therefore, treating the observed number of active bidders as the number of potential bidders can lead to incorrect inference. The next theorem studies identification when the number of potential bidders is unobserved but the number of active bidders is observed. \\

\begin{theorem}
Suppose the number of potential bidders $N$ is independent of bidder values, finite, and unobserved by the econometrician. Assume the number of active bidders is observed. Then $\alpha^*$ and $V(\alpha)$ for $\alpha\geq\alpha^*$ are point identified in first-price auctions but not in second-price auctions. \\
\end{theorem}

Theorem 4 highlights that the observability of active bidders provides different amounts of information across auction formats. In first-price auctions, the observed number of active bidders contains information about the distribution of potential competition and can be combined with transaction prices to recover the screening threshold. In contrast, in second-price auctions, even observing active bidders does not provide sufficient information to separate the distribution of potential bidders from the screening mechanism. As a result, the bidder value distribution remains unidentified.

The non-identification result in second-price auctions is particularly strong. Even if the full vector of bids were observed, the bidder value distribution would generally remain unidentified under this information structure. However, additional information from invalid auctions can partially restore identification in second-price auctions, which is shown in the next theorem. \\

\begin{theorem}
Suppose the number of potential bidders $N$ is independent of bidder values, strictly positive, finite, and unobserved by the econometrician. Assume the number of active bidders and the number of invalid auctions are observed. Then $\alpha^*$ and $V(\alpha)$ for $\alpha \geq \alpha^*$ are set identified in second-price auctions. The identified set contains finitely many elements and is given in the appendix.\\
\end{theorem}

Theorem 5 demonstrates that information on unsuccessful auctions provides additional identifying restrictions even when the number of potential bidders is unobserved. The frequency of invalid auctions reveals the probability that all potential bidders are screened out, while the observed number of active bidders provides information on bidder participation conditional on a sale. Together, these two sources of information restrict the possible values of the screening threshold, yielding a finite identified set for $(\alpha^*,V(\alpha))$.

\subsection{Summary}

The results in this section provide a complete characterization of identification under the information structures considered when transaction prices are truncated by a binding reserve price. Although the identification results differ across auction formats and information structures, they all reflect a common principle: truncation creates an unknown screening threshold, and identification hinges on whether the available data contain sufficient information to recover this threshold.

Several general lessons emerge. When the number of potential bidders is fixed, the observed transaction price distribution alone is generally insufficient to recover the screening threshold. Consequently, identification relies on additional information, such as the mass point at the reserve price in second-price auctions or the frequency of no-sale auctions in first-price auctions. When the number of potential bidders varies across auctions, exogenous variation in competition provides additional identifying restrictions by linking auctions through a common screening threshold. Whether these restrictions are sufficient ultimately depends on the information available to the econometrician, in particular, whether the number of potential bidders or only the number of active bidders is observed.

The next section extends the analysis to auctions with entry costs. Although the source of truncation differs, many of the information structures considered in this section continue to play analogous roles. The identifying arguments, however, differ because the screening threshold generally varies across auctions with different numbers of potential bidders.

\section{Identification with Entry Costs}

This section extends the analysis to auctions with entry costs. As in the binding reserve-price environment, transaction prices are observed only when at least one bidder participates, giving rise to truncated transaction price data. The key difference is that truncation now results from endogenous participation decisions rather than an exogenously imposed reserve price, and an invariant entry cost does not generally imply an invariant screening threshold. Consequently, although the information structures considered here are identical to those in Section 4, the identifying arguments differ in important ways. The following assumption characterizes the entry-cost environment considered in this section. \\

\textbf{Assumption E:}

\textit{(i) The support of $F(\cdot)$ is $[\underline{v},\bar{v}]$, where $0\leq \underline{v}<\bar{v}<\infty$.}

\textit{(ii) $F(\cdot)$ admits a continuous and strictly positive density $f(\cdot)$ on its support.}

\textit{(iii) The reserve price $R$ is constant across auctions and satisfies $R<\bar{v}$.}

\textit{(iv) The entry cost $c$ is constant across auctions and bidders, with $c\in(0,\bar{v}-R)$.}

\textit{(v) All bidders play the symmetric and strictly increasing equilibrium strategy.} \\

Assumptions E(i), E(ii), and E(v) are identical to Assumptions R(i), R(ii), and R(iv), respectively. Assumption E(iii) allows the reserve price to lie below the lower bound of bidder values, reflecting the fact that truncation may arise solely from entry costs even in the absence of a binding reserve price. Assumption E(iv) is the key identifying assumption in this section. Unlike the reserve-price environment, an invariant entry cost does not imply an invariant screening threshold. Because bidders decide whether to enter strategically, the equilibrium entry threshold depends jointly on the entry cost and the number of potential bidders. Consequently, the screening threshold generally varies across auctions even when the entry cost is constant. Instead of exploiting an invariant screening threshold, the identification analysis relies on the equilibrium relationship between entry costs, participation decisions, and transaction prices. Finally, the restriction $c>0$ ensures that the equilibrium bidding function has a strictly positive derivative at the entry threshold, which plays an important role in the identification results for first-price auctions.

Before presenting the identification results, it is useful to clarify the object of identification in the presence of entry costs. Let $\alpha_N^*$ denote the equilibrium entry threshold when there are $N$ potential bidders. Unlike the reserve-price environment, where the screening threshold is invariant across auctions, the entry threshold generally depends on $N$. Consequently, the identification problem is no longer characterized by a single unknown screening threshold, but by the collection of equilibrium entry thresholds together with the bidder value distribution.

Lemma 1 below characterizes the equilibrium entry and bidding strategies. As in Section 4, the identification analysis focuses on recovering the bidder value distribution above the relevant entry threshold. Since the equilibrium thresholds are determined jointly by the entry cost and the bidder value distribution, identification amounts to recovering the entry cost together with the bidder value distribution from the observed transaction prices under the alternative information structures considered below. \\

\begin{lemma}
Suppose a bidder has type $\alpha$. In first-price auctions, the bidder's equilibrium strategy is not to enter if $\alpha<\alpha_N^*$; enter and bid $b(\alpha)$ if $\alpha \geq \alpha_N^*$, where
\[
b(\alpha) = V(\alpha) - \frac{1}{\alpha^{N-1}} \left[ c+\int_{\alpha_N^*}^\alpha V'(t) t^{N-1} d t \right],
\]
and $\alpha_N^*$ is the unique solution to
\[
\left[ V(\alpha_N^*)-R \right] (\alpha_N^*)^{N-1} - c = 0.
\]
In second-price auctions, the bidder's equilibrium strategy is not to enter if $\alpha<\alpha_N^*$; enter and bid $V(\alpha)$ if $\alpha \geq \alpha_N^*$, where $\alpha_N^*$ is again the unique solution to
\[
\left[ V(\alpha_N^*)-R \right] (\alpha_N^*)^{N-1} - c = 0.
\] \\
\end{lemma}

Lemma 1 establishes three properties of the equilibrium that are central to the identification analysis. First, the equilibrium entry threshold depends jointly on the entry cost and the number of potential bidders, and is increasing in each of them. Consequently, unlike the binding reserve-price environment, the screening threshold is generally not invariant across auctions with different numbers of potential bidders. This distinction is the key reason why the identifying arguments in this section differ from those in Section 4.

Second, the equilibrium entry threshold is identical in first- and second-price auctions. Therefore, the differences in identification established below arise entirely from the bidding stage rather than from bidders' participation decisions.

Finally, differentiating the equilibrium bidding function in first-price auctions yields
\[
b'(\alpha)
=
\frac{N-1}{\alpha^N}
\left[
c+\int_{\alpha_N^*}^{\alpha}V'(t)t^{N-1}dt
\right].
\]
In particular,
\[
b'(\alpha_N^*)
=
\frac{N-1}{(\alpha_N^*)^N}c>0,
\]
where the inequality follows from Assumption E(iv). By contrast, the corresponding derivative satisfies $b'(\alpha^*)=0$ in the reserve-price environment. As shown in Theorem 7, this distinction plays a crucial role in the identification of first-price auctions.

Despite the different truncation mechanism, the identification analysis follows the same sequence as in Section 4. I begin with the benchmark case in which the number of potential bidders is fixed and observed. I then consider auctions with varying numbers of potential bidders, where cross-auction variation provides additional identifying information. As will be seen, although the information structures are identical to those in the reserve-price environment, the identification results differ because the screening threshold varies across auctions.

\subsection{Fixed Number of Potential Bidders}

I first consider the benchmark case where all auctions have the same number of potential bidders, which is known to the econometrician. As in Section 4, this setting isolates the role of truncation by removing variation in the level of competition across auctions. The identification problem therefore centers on whether the observed transaction prices contain sufficient information to recover the entry cost and the bidder value distribution. \\

\begin{theorem}
Suppose all auctions share a common number of potential bidders $N\geq2$, which is known by the econometrician. Then $c$, $\alpha_N^*$, and $V(\alpha)$ for $\alpha\geq\alpha_N^*$ are point identified in second-price auctions but not in first-price auctions. \\
\end{theorem}

Theorem 6 is the analogue of Theorem 1 for the entry-cost environment. As in the reserve-price case, identification hinges on recovering the screening threshold. The key difference is that the threshold is now determined endogenously through bidders' entry decisions rather than directly by the reserve price. In second-price auctions, the observed transaction price distribution again contains a mass point, allowing the econometrician to recover the entry threshold and, consequently, the entry cost and the bidder value distribution above the threshold.

In contrast, first-price auctions do not exhibit such a mass point. The observed transaction prices therefore reveal only the equilibrium bids of entering bidders, making it impossible to separately recover the entry threshold and the entry cost. Consequently, neither the entry cost nor the bidder value distribution is point identified from transaction prices alone.

The next theorem shows that point identification in first-price auctions can be restored when the entry cost is observed. \\

\begin{theorem}
Suppose all auctions share a common number of potential bidders $N\geq2$, which is known by the econometrician. If the entry cost $c$ is observed, then $\alpha_N^*$ and $V(\alpha)$ for $\alpha\geq\alpha_N^*$ are point identified in first-price auctions. \\
\end{theorem}

Theorem 7 shows that the failure of identification in first-price auctions is driven entirely by the inability to recover the entry threshold. Once the entry cost is observed, the equilibrium entry condition uniquely determines the threshold, thereby restoring identification of the bidder value distribution.

A notable difference from the reserve-price environment is that the identifying argument relies critically on the assumption that $c>0$. By Lemma 1,
\[
b'(\alpha_N^*)
=
\frac{N-1}{(\alpha_N^*)^N}c,
\]
which is strictly positive under Assumption E(iv). If $c=0$, then $b'(\alpha_N^*)=0$, and the identifying equation used to recover the entry threshold degenerates. Thus, unlike the reserve-price environment studied in Section 4, a strictly positive entry cost is essential for identification in first-price auctions.

\subsection{Varying Number of Potential Bidders}

I now consider the case where the number of potential bidders varies across auctions. As in Section 4, exogenous variation in the level of competition provides additional identifying information. The key difference from the reserve-price environment, however, is that the equilibrium entry threshold now varies with the number of potential bidders. Consequently, variation in competition no longer links auctions through a common screening threshold. Instead, identification exploits the equilibrium relationship between the entry cost, the sequence of entry thresholds, and the observed transaction prices. I first consider the case where the number of potential bidders is observed. \\

\begin{theorem}
Suppose the number of potential bidders $N$ is observed and takes at least two different values greater than one. Assume the bidder value distribution is invariant across auctions with different numbers of potential bidders. Then $c$ and $V(\alpha)$ for $\alpha$ above some threshold are set identified in first-price auctions. The identified set contains finitely many elements and is given in the appendix. \\
\end{theorem}

Theorem 8 is the counterpart of Theorem 3 in the entry-cost environment. Unlike the reserve-price case, where variation in the number of potential bidders leads to point identification, variation in competition here generally yields only finite set identification. The reason is that the screening threshold itself changes with the number of potential bidders. As a result, cross-auction variation generates a system of equilibrium restrictions rather than a common screening threshold that directly links different auctions. Nevertheless, these restrictions substantially narrow the identified set and may deliver point identification when sufficiently rich variation in the number of potential bidders is available.

By Theorem 6, second-price auctions are already point identified when the number of potential bidders is fixed and observed. Consequently, variation in the number of potential bidders provides additional restrictions beyond those required for identification, rendering the model overidentified. This makes the assumption that the bidder value distribution is invariant across auctions testable.

In many empirical settings, however, the econometrician observes only the number of active bidders rather than the number of potential bidders. The next theorem considers this information structure. \\

\begin{theorem}
Suppose the number of potential bidders $N$ is independent of bidder values, finite, and unobserved by the econometrician. Assume the number of active bidders is observed. Then $c$ and $V(\alpha)$ for $\alpha$ above some threshold are point identified in first-price auctions but not in second-price auctions. \\
\end{theorem}

Theorem 9 reveals a striking reversal relative to the case where the number of potential bidders is observed. When only the number of active bidders is available, first-price auctions become more informative than second-price auctions. The intuition lies in the equilibrium bidding behavior. In first-price auctions, bidders adjust their bids according to the number of potential competitors. Although the econometrician does not observe the number of potential bidders, this variation is reflected in the observed transaction price distribution and can be exploited for identification.

By contrast, truthful bidding remains optimal in second-price auctions. Conditional on entry, bids no longer depend on the number of potential bidders, so variation in competition affects the observed data only through entry decisions. Consequently, transaction prices and the number of active bidders do not contain sufficient information to recover the entry cost and the bidder value distribution.

The next theorem shows that point identification in second-price auctions can be restored if the entry cost is observed. \\

\begin{theorem}
Suppose the number of potential bidders $N$ is independent of bidder values, unobserved by the econometrician, and has support $\{2,3,\ldots,\bar N\}$. Assume the number of active bidders and the entry cost are observed. Then the distribution of the number of potential bidders and $V(\alpha)$ for $\alpha$ above some threshold are point identified in second-price auctions. \\
\end{theorem}

Theorem 10 demonstrates that observing the entry cost provides exactly the additional information needed to recover the distribution of potential bidders. Once the entry thresholds are identified through the equilibrium entry condition, the remaining model primitives can be recovered from the joint distribution of transaction prices and active bidders. Thus, although second-price auctions are not point identified under the information structure of Theorem 9, they become point identified once the entry cost is observed.

\subsection{Summary}

The results in this section provide a complete characterization of identification under the information structures considered when transaction prices are truncated by entry costs. As in the binding reserve-price environment, identification depends jointly on the available information regarding bidder participation, the variation in the number of potential bidders, and the auction format. The key distinction is that the screening threshold is now determined endogenously through bidders' entry decisions and therefore generally varies with the number of potential bidders.

In the entry cost environment, when the number of potential bidders is fixed, identification again hinges on recovering the screening threshold. In second-price auctions, this can be achieved from the transaction price distribution, whereas first-price auctions additionally require information on the entry cost. When the number of potential bidders varies across auctions, exogenous variation in competition continues to provide additional identifying restrictions. Unlike the reserve-price environment, however, this variation no longer links auctions through a common screening threshold. Instead, identification exploits the system of equilibrium restrictions generated by the entry model, leading to different identification results under the alternative information structures.

Comparing Sections 4 and 5 reveals a common identification framework for auctions with truncated transaction prices. Although binding reserve prices and entry costs generate truncation through different economic mechanisms, the information available to the econometrician plays a remarkably similar role in determining identification. This comparison highlights that identification is governed jointly by the source of truncation and the available information on bidder participation, providing a unified perspective on identification in auction models with truncated transaction prices.

\section{Conclusion}

This paper studies identification in auction models when transaction price data are truncated due to either a binding reserve price or endogenous entry. Although both sources of truncation are prevalent in practice, they have received relatively little attention in the identification literature. This paper develops a unified framework for studying identification under these two truncation mechanisms and characterizes the information required to recover the bidder value distribution under a range of auction formats and information structures.

A central insight of the paper is that identification hinges on recovering the screening threshold that determines which bidders are observed in the data. In the binding reserve-price environment, the screening threshold is determined directly by the seller's reserve price and remains invariant across auctions under the maintained assumptions. By contrast, in the entry-cost environment, the screening threshold is determined endogenously through bidders' participation decisions and generally varies with the number of potential bidders. This distinction leads to different identifying arguments, despite the similar information structures considered in the two environments.

The analysis reveals that identification depends jointly on the source of truncation, the auction format, and the information available to the econometrician. When the number of potential bidders is fixed, the screening threshold can be recovered directly from the observed transaction price in second-price auctions, whereas first-price auctions require additional information, such as the frequency of no-sale auctions or the entry cost. When the number of potential bidders varies across auctions and is observed, exogenous variation in competition provides additional identifying restrictions that often yield point identification. By contrast, when the number of potential bidders is unobserved but the number of active bidders is observed, the identification pattern is reversed: first-price auctions remain point identified, whereas second-price auctions generally do not. More broadly, the results demonstrate that bidder participation contains valuable identifying information and that seemingly limited data—such as the frequency of no-sale auctions or the number of active bidders—can substantially strengthen identification.

While the paper focuses on symmetric independent private-value auctions, the framework suggests several promising directions for future research. An important extension is to study identification under more general environments, including asymmetric bidders, affiliated or interdependent values, and unobserved auction heterogeneity. Another natural direction is to develop estimation methods based on the constructive identification results established here. Such methods would make it possible to estimate auction models using transaction price data that are truncated by reserve prices or endogenous entry, thereby extending the applicability of structural auction models to a broader range of empirical settings.

\newpage

\section*{Appendix}

\subsection*{Proof of Lemma 1}

First consider first-price auctions. Guess that there exists an equilibrium where bidders whose private values $\alpha_i<\alpha_N^*$ do not enter, and bidders whose private values $\alpha_i \geq \alpha_N^*$ bid $b(\alpha_i)$ which is strictly increasing in $\alpha_i$ with $b(\alpha_N^*)=R$. Now I derive $b(\alpha)$ and show it is indeed an equilibrium. First, for bidders with $\alpha_i=\alpha_N^*$ to be indifferent with bidding $R$ and not bidding, there must be
\[
\left[ V(\alpha_N^*)-R \right] \cdot (\alpha_N^*)^{N-1} = c.
\]
Since LHS strictly increases with $\alpha^*$ and $0<c<V(1)-R$, there exists a unique solution for $\alpha_N^*$. For $b(\alpha)$ to be an equilibrium, it must be optimal for a bidder to bid $b(\alpha)$ when his true type is $\alpha$. Note that by bidding $b(\beta)$ for some $\beta \geq \alpha_N^*$, this bidder's expected payoff is
\[
\pi(\beta;\alpha) = [V(\alpha)-b(\beta)] \beta^{N-1} - c.
\]
Taking derivative, we have
\[
\frac{\partial \pi}{\partial \beta} = [V(\alpha)-b(\beta)] (N-1) \beta^{N-2} - b'(\beta) \beta^{N-1}.
\]
The FOC then gives
\[
\left. \frac{\partial \pi}{\partial \beta} \right|_{\beta=\alpha} = [V(\alpha)-b(\alpha)] (N-1) \alpha^{N-2} - b'(\alpha) \alpha^{N-1} = 0,
\]
which implies
\[
\frac{d}{d \alpha} \left[b(\alpha) \alpha^{N-1}\right] = V(\alpha) \cdot \frac{d}{d \alpha} \left(\alpha^{N-1}\right).
\]
Since $b(\alpha_N^*)=R$ and $[V(\alpha_N^*)-R] (\alpha_N^*)^{N-1} = c$, we then have
\[
\begin{aligned}
b(\alpha) &= \frac{1}{\alpha^{N-1}} \int_{\alpha_N^*}^{\alpha} V(t) d t^{N-1} \\
&= V(\alpha) - \frac{1}{\alpha^{N-1}} \left[ c + \int_{\alpha_N^*}^{\alpha} V'(t)t^{N-1} dt \right].
\end{aligned}
\]
Now we verify this bidding strategy is indeed an equilibrium. First, note that
\[
\pi(\alpha;\alpha) = \int_{\alpha_N^*}^{\alpha} V'(t)t^{N-1} dt > 0 \ \ \ \ \forall \alpha>\alpha_N^*.
\]
Thus when $\alpha>\alpha_N^*$, bidding $b(\alpha)$ is better than not bidding. Second, since
\[
\pi(\beta;\alpha) = \beta^{N-1}[V(\alpha)-V(\beta)] + \int_{\alpha_N^*}^{\beta} V'(t)t^{N-1} dt,
\]
we have
\[
\Delta(\beta;\alpha) \equiv \pi(\beta;\alpha)-\pi(\alpha;\alpha) =  \beta^{N-1}[V(\alpha)-V(\beta)] + \int_{\alpha}^{\beta} V'(t)t^{N-1} dt.
\]
Note that
\[
\frac{\partial \Delta}{\partial \beta} = (N-1)\beta^{N-2}[V(\alpha)-V(\beta)],
\]
thus
\[
\begin{aligned}
&\frac{\partial \Delta}{\partial \beta}>0 \Leftrightarrow \beta<\alpha, \\
&\frac{\partial \Delta}{\partial \beta}<0 \Leftrightarrow \beta>\alpha, 
\end{aligned}
\]
which implies
\[
\Delta(\beta;\alpha) \leq \Delta(\alpha;\alpha) = 0.
\]
Therefore, when $\alpha>\alpha_N^*$, bidding $b(\alpha)$ is indeed optimal given that all other bidders play the same strategy. Finally, we need to show when $\alpha<\alpha_N^*$, it is optimal not to enter. Suppose this bidder bids $b(\beta)$ for some $\beta \geq \alpha_N^*$, then his expected payoff
\[
\pi(\beta;\alpha) < \pi(\beta;\alpha_N^*) = \beta^{N-1}[V(\alpha_N^*)-V(\beta)] \leq 0,
\]
which completes the proof. 

Now consider second-price auctions. If a bidder decides to enter a second-price auction, then it is always a weakly dominant strategy to bid his true value. Therefore, it suffices to determine the entry decision. Suppose only bidders with types $\alpha_i \geq \alpha_N^*$ enter, then the expected payoff for bidders with types $\alpha_i=\alpha_N^*$ equals 
\[
\pi(\alpha_N^*) = \left[ V(\alpha_N^*)-R \right] (\alpha_N^*)^{N-1} - c.
\]
In equilibrium, there should be $\pi(\alpha_N^*)=0$, which implies $\alpha_N^*$ is defined by
\[
\left[ V(\alpha_N^*)-R \right] (\alpha_N^*)^{N-1} = c.
\]
By similar arguments as before, it can be checked that this is indeed an equilibrium. \hfill $\blacksquare$ \\

\subsection*{Proof of Theorem 1}

To begin with, I show $\alpha^*$ and $V(\alpha)$ for $\alpha \geq \alpha^*$ are not identified in first-price auctions. Denote the quantile function of the observed transaction price by $T(\cdot)$. Suppose $T(\cdot)$ is generated by $\alpha_1>0$ and $V_1(\cdot)$. Denote the bidding function for $\alpha_1$ and $V_1(\cdot)$ by $b_1(\cdot)$. Recall that 
\[
\frac{d}{d \alpha} \left[b_1(\alpha) \alpha^{N-1}\right] = V_1(\alpha) \cdot \frac{d}{d \alpha} \left(\alpha^{N-1}\right),
\]
which is equivalent to
\[
b_1(\alpha) + \frac{\alpha}{N-1} b_1'(\alpha) = V_1(\alpha). \label{2}
\]
Since $b_1(\alpha_1)=V_1(\alpha_1)$, there must be $b_1'(\alpha_1)=0$. Also note that
\[
\begin{aligned}
\mathbb{P}\left(T_l<b_1(\alpha)\right) 
& =\mathbb{P}\left(b_1\left(\alpha_i^{N:N}\right)<b_1(\alpha) \mid \alpha_i^{N:N} \geq \alpha_1\right) \\
& =\frac{\mathbb{P}\left(\alpha_1 \leq \alpha_i^{N: N}<\alpha\right)}{\mathbb{P}\left(\alpha_i^{N: N} \geq \alpha_1\right)} \\
& =\frac{\alpha^N-\alpha_1^N}{1-\alpha_1^N}, 
\end{aligned}
\]
which implies
\[
b_1(\alpha) = T \left( \frac{\alpha^N-\alpha_1^N}{1-\alpha_1^N} \right).
\]
Taking derivative, we have
\[
b_1'(\alpha) = \frac{N\alpha^{N-1}}{1-\alpha_1^N} T' \left( \frac{\alpha^N-\alpha_1^N}{1-\alpha_1^N} \right),
\]
Let $\alpha=\alpha_1$, we get
\[
T'(0) = \frac{1-\alpha_1^N}{N\alpha_1^{N-1}} b_1'(\alpha_1) = 0.
\]
Pick $\alpha_2 \in (0,1)$ arbitrarily and let
\[
\begin{aligned}
&b_2(\alpha) = T \left( \frac{\alpha^N-\alpha_2^N}{1-\alpha_2^N} \right)  \ \ \forall \alpha \geq \alpha_2, \\
&V_2(\alpha) = b_2(\alpha) + \frac{\alpha}{N-1}b_2'(\alpha). \\
\end{aligned}
\]
Now I show $\alpha_2$ and $V_2(\cdot)$ also generate $T(\cdot)$. To show $V_2(\cdot)$ generate $b_2(\cdot)$, note that $b_2'(\alpha_2)=T'(0)=0$, which implies $b_2(\alpha_2)=V_2(\alpha_2)$. Therefore, $b_2(\cdot)$ satisfies both the FOC and the boundary condition, and thus is the equilibrium bidding function for $V_2(\cdot)$. To show $\alpha_2$ and $b_2(\cdot)$ generate $T(\cdot)$, note that for all $\alpha \geq \alpha_2$,
\[
\begin{aligned}
\mathbb{P}\left(T_l<b_2(\alpha)\right) 
& =\mathbb{P}\left(b_2\left(\alpha_i^{N:N}\right)<b_2(\alpha) \mid \alpha_i^{N:N} \geq \alpha_2\right) \\
& =\frac{\mathbb{P}\left(\alpha_2 \leq \alpha_i^{N: N}<\alpha\right)}{\mathbb{P}\left(\alpha_i^{N: N} \geq \alpha_2\right)} \\
& =\frac{\alpha^N-\alpha_2^N}{1-\alpha_2^N}, 
\end{aligned}
\]
which implies
\[
b_2(\alpha) = T \left( \frac{\alpha^N-\alpha_2^N}{1-\alpha_2^N} \right).
\]
Since the choice of $\alpha_2$ is arbitrary, $\alpha^*$ and $V(\alpha)$ for $\alpha \geq \alpha^*$ are not identified in first-price auctions.

Now I show $\alpha^*$ and $V(\alpha)$ for $\alpha \geq \alpha^*$ can be identified in second-price auctions. Denote the number of active bidders by $N^{\text{act}}$, then we have $\min\{T_l\} \xrightarrow{p} R$ and
\[
\begin{aligned}
\frac{1}{L} \sum_{l=1}^L \mathbb{1}(T_l=R) &\xrightarrow{p} \mathbb{P}(N^+=1) \\
&= \mathbb{P}(N^{\text{act}}=1 \mid N^{\text{act}} \geq 1) \\
&= \frac{N(1-\alpha^*)(\alpha^*)^{N-1}}{1-(\alpha^*)^N}.
\end{aligned}
\]
Let
\[
h(\alpha) = \frac{(1-\alpha)\alpha^{N-1}}{1-\alpha^N}.
\]
To show $\alpha^*$ is point identified, it suffices to show $h(\alpha)$ is strictly increasing in $(0,1)$. Taking derivative, we get
\[
\begin{aligned}
h'(\alpha) &= \frac{1}{(1-\alpha^N)^2} \left\{ \left[(N-1)\alpha^{N-2}-N\alpha^{N-1}\right](1-\alpha^N) - (\alpha^{N-1}-\alpha^N)(-N\alpha^{N-1}) \right\} \\
&= \frac{\alpha^{N-2}}{(1-\alpha^N)^2} \left[ (N-1)-N\alpha-(N-1)\alpha^N+N\alpha^{N+1}+N\alpha^N-N\alpha^{N+1} \right]  \\
&= \frac{\alpha^{N-2}}{(1-\alpha^N)^2} \left[ (N-1)-N\alpha+\alpha^N \right].
\end{aligned}
\]
Let
\[
g(\alpha) = (N-1)-N\alpha+\alpha^N,
\]
we have
\[
g'(\alpha) = -N+N\alpha^{N-1} < 0.
\]
which implies $g(\alpha)>g(1)=0$, and thus $h'(\alpha)>0$ for $\alpha \in (0,1)$. Now focus on auctions with $T_l>R$ and denote the quantile function of transaction price in these auctions by $T(\cdot)$. Then we have
\[
\begin{aligned}
\mathbb{P}(T_l<V(\alpha) \mid T_l>R) &= \mathbb{P}(v^{N-1:N}<V(\alpha) \mid v^{N-1:N}>V(\alpha^*)) \\
&= \frac{\mathbb{P}(v^{N-1:N}<V(\alpha))-\mathbb{P}(v^{N-1:N} \leq V(\alpha^*))}{1-\mathbb{P}(v^{N-1:N} \leq V(\alpha^*))} \\
&= \frac{\phi(\alpha)-\phi(\alpha^*)}{1-\phi(\alpha^*)},
\end{aligned}
\]
where
\[
\phi(\alpha) = N\alpha^{N-1}-(N-1)\alpha^N.
\]
Therefore, for $\alpha \geq \alpha^*$, $V(\alpha)$ can be identified by
\[
V(\alpha) = T \left( \frac{\phi(\alpha)-\phi(\alpha^*)}{1-\phi(\alpha^*)} \right),
\]
which completes the proof. \hfill $\blacksquare$ \\

\subsection*{Proof of Theorem 2}

If $L^{\text{invalid}}$ is observed, note that
\[
\frac{L^{\text{invalid}}}{L+L^{\text{invalid}}} \xrightarrow{p} (\alpha^*)^N,
\]
thus $\alpha^*$ is identified. Let $T(\cdot)$ denote the quantile function of $T_l$, then by similar arguments in Theorem 1, $V(\cdot)$ for all $\alpha \geq \alpha^*$ can be point identified by
\[
\begin{aligned}
V(\alpha) &= b(\alpha) + \frac{\alpha}{N-1} b'(\alpha) \\
&= T \left( \frac{\alpha^N-(\alpha^*)^N}{1-(\alpha^*)^N} \right) + \frac{N}{N-1} \frac{\alpha^N}{1-(\alpha^*)^N} T' \left( \frac{\alpha^N-(\alpha^*)^N}{1-(\alpha^*)^N} \right),
\end{aligned}
\]
which completes the proof. \hfill $\blacksquare$ \\

\subsection*{Proof of Theorem 3}

It suffices to prove identification in first-price auctions. Pick $N_1$, $N_2$ from the known $N$'s such that $N_1>N_2>1$. Denote the corresponding quantile functions of the transaction price by $T_1(\cdot)$ and $T_2(\cdot)$. From the previous part, we know that the optimal screening level does not change with $N$, thus $\alpha_1^*=\alpha_2^*=\alpha^*$. By similar arguments as in Theorem 1, we have
\[
\begin{aligned}
V(\alpha) &= b_1(\alpha) + \frac{\alpha}{N_1-1}b_1'(\alpha) \\
&= T_1 \left( \frac{\alpha^{N_1}-(\alpha^*)^{N_1}}{1-(\alpha^*)^{N_1}} \right) + \frac{N_1}{N_1-1} \frac{\alpha^{N_1}}{1-(\alpha^*)^{N_1}} T_1' \left( \frac{\alpha^{N_1}-(\alpha^*)^{N_1}}{1-(\alpha^*)^{N_1}} \right),
\end{aligned}
\]
and
\[
\begin{aligned}
V(\alpha) &= b_2(\alpha) + \frac{\alpha}{N_2-1}b_2'(\alpha) \\
&= T_2 \left( \frac{\alpha^{N_2}-(\alpha^*)^{N_2}}{1-(\alpha^*)^{N_2}} \right) + \frac{N_2}{N_2-1} \frac{\alpha^{N_2}}{1-(\alpha^*)^{N_2}} T_2' \left( \frac{\alpha^{N_2}-(\alpha^*)^{N_2}}{1-(\alpha^*)^{N_2}} \right).
\end{aligned}
\]
If $T_1\left(\alpha^{N_1}\right) + \frac{N_1\alpha^{N_1}}{N_1-1} T_1'\left(\alpha^{N_1}\right) = T_2\left(\alpha^{N_2}\right) + \frac{N_2\alpha^{N_2}}{N_2-1} T_2'\left(\alpha^{N_2}\right)$, then $\alpha^*=0$ and $V(\alpha)$ for $\alpha \geq \alpha^*$ can be identified by
\[
V(\alpha) = T_1\left(\alpha^{N_1}\right) + \frac{N_1\alpha^{N_1}}{N_1-1} T_1'\left(\alpha^{N_1}\right).
\]
Otherwise, since $V(\alpha^*)=T_1(0)=T_2(0)$, there must be $T_1'(0)=T_2'(0)=0$.

Before proceeding, note that $T_1''(0)$ and $T_2''(0)$ are well defined under Assumption R(ii). In particular, for $j=1,2$,
\[
b_j(\alpha)
=
V(\alpha)
-
\frac{1}{\alpha^{N_j-1}}
\int_{\alpha^*}^{\alpha}V'(t)t^{N_j-1}dt,
\]
which implies
\[
b_j'(\alpha)
=
\frac{N_j-1}{\alpha^{N_j}}
\int_{\alpha^*}^{\alpha}V'(t)t^{N_j-1}dt.
\]
Since $V'(\cdot)$ is continuous under Assumption R(ii), $b_j(\cdot)$ is twice continuously differentiable. Moreover,
\[
b_j(\alpha)
=
T_j\left(
\frac{\alpha^{N_j}-(\alpha^*)^{N_j}}
{1-(\alpha^*)^{N_j}}
\right),
\]
and the mapping
\[
\alpha
\mapsto
\frac{\alpha^{N_j}-(\alpha^*)^{N_j}}
{1-(\alpha^*)^{N_j}}
\]
is twice continuously differentiable with a strictly positive derivative at $\alpha^*>0$. Hence its inverse is twice continuously differentiable in a neighborhood of zero, implying that $T_j(\cdot)$ is twice continuously differentiable at zero.

Taking derivatives and letting $\alpha=\alpha^*$, we then have
\[
\begin{aligned}
V'(\alpha^*) &= \frac{N_1}{N_1-1} \frac{(\alpha^*)^{N_1}}{1-(\alpha^*)^{N_1}} \frac{N_1 (\alpha^*)^{N_1-1}}{1-(\alpha^*)^{N_1}} T_1''(0) \\
&= \frac{N_2}{N_2-1} \frac{(\alpha^*)^{N_2}}{1-(\alpha^*)^{N_2}} \frac{N_2 (\alpha^*)^{N_2-1}}{1-(\alpha^*)^{N_2}} T_2''(0),
\end{aligned}
\]
which implies
\[
\frac{N_1^2}{N_1-1} \left[ \frac{(\alpha^*)^{N_1}}{1-(\alpha^*)^{N_1}} \right]^2 T_1''(0) = \frac{N_2^2}{N_2-1} \left[ \frac{(\alpha^*)^{N_2}}{1-(\alpha^*)^{N_2}} \right]^2 T_2''(0) \neq 0.
\]
Let
\[
f(\alpha) = \alpha^{N_1-N_2} \cdot \frac{1-\alpha^{N_2}}{1-\alpha^{N_1}}.
\]
Since $\alpha \in (0,1)$, $f(\alpha)>0$, and thus
\[
f(\alpha^*) = \left[ \frac{N_2^2(N_1-1)}{N_1^2(N_2-1)} \frac{T_2''(0)}{T_1''(0)} \right]^{\frac{1}{2}}.
\]
Now we show $f$ is strictly monotone on $(0,1)$. Note that
\[
f'(\alpha) = \frac{\alpha^{N_1-N_2-1}}{(1-\alpha^{N_1})^2} \left[ N_1(1-\alpha^{N_2})-N_2(1-\alpha^{N_1}) \right],
\]
thus it suffices to show $N_1(1-\alpha^{N_2})-N_2(1-\alpha^{N_1})$ is strictly positive. Let
\[
g(\alpha) = N_1(1-\alpha^{N_2})-N_2(1-\alpha^{N_1}),
\]
we have
\[
g'(\alpha) = N_1N_2\alpha^{N_2-1}(\alpha^{N_1-N_2}-1) < 0.
\]
Therefore, $g(\alpha)>g(1)=0$ for all $\alpha \in (0,1)$, which implies $f'(\alpha)>0$, and thus $\alpha^*$ is identified. Then $V(\alpha)$ for $\alpha \geq \alpha^*$ can be identified by
\[
V(\alpha) = T_1 \left( \frac{\alpha^{N_1}-(\alpha^*)^{N_1}}{1-(\alpha^*)^{N_1}} \right) + \frac{N_1}{N_1-1} \frac{\alpha^{N_1}}{1-(\alpha^*)^{N_1}} T_1' \left( \frac{\alpha^{N_1}-(\alpha^*)^{N_1}}{1-(\alpha^*)^{N_1}} \right),
\]
which completes the proof. \hfill $\blacksquare$ \\

\subsection*{Proof of Theorem 4}

First consider first-price auctions. Denote the quantile function of the transaction price conditional on $N^+=k$ by $T_k(\cdot)$ and the bidder’s bidding strategy when $N=k$ by $b_k(\cdot)$. Denote the maximum number of bidders by $\bar{N}$, then $\bar{N}$ can be identified by $\max\{N_l^+\}$. First focus on auctions with $N_l^+=\bar{N}$. Since there must be $N_l=\bar{N}$ when $N_l^+=\bar{N}$, for all $\alpha>\alpha^*$, we have
\[
\begin{aligned}
\mathbb{P} \left( T_l<b_{\bar{N}}(\alpha) \mid N^+=\bar{N} \right) &= \mathbb{P}\left(T_l<b_{\bar{N}}(\alpha) \mid N^+=\bar{N}, N=\bar{N}\right) \\
&= \mathbb{P} \left( b_{\bar{N}}(\alpha_i^{\bar{N}:\bar{N}})<b_{\bar{N}}(\alpha) \mid \alpha_i>\alpha^* \ \ \forall i \leq \bar{N} \right) \\
&= [\mathbb{P} \left( \alpha_i<\alpha \mid \alpha_i>\alpha^* \right)]^{\bar{N}}\\
&= \left(\frac{\alpha-\alpha^*}{1-\alpha^*}\right)^{\bar{N}},
\end{aligned}
\]
which implies
\[
b_{\bar{N}}(\alpha) = T_{\bar{N}}\left(\left(\frac{\alpha-\alpha^*}{1-\alpha^*}\right)^{\bar{N}}\right) .
\]
Taking derivative, we get
\[
b_{\bar{N}}'(\alpha) = T_{\bar{N}}'\left(\left(\frac{\alpha-\alpha^*}{1-\alpha^*}\right)^{\bar{N}}\right) \cdot \bar{N}\left(\frac{\alpha-\alpha^*}{1-\alpha^*}\right)^{\bar{N}-1} \frac{1}{1-\alpha^*}.
\]
Now focus on auctions with $N_l^+=\bar{N}-1$. Note that $b_{\bar{N}}(1)>b_{\bar{N}-1}(1)$, thus there exists $\beta \in (0,1)$ such that $b_{\bar{N}}(\beta)=b_{\bar{N}-1}(1)$. Let $p=\mathbb{P}\left(N=\bar{N} \mid N^+=\bar{N}-1\right)$, then for $\alpha>\beta$, we have
\[
\begin{aligned}
\mathbb{P} \left( T_l<b_{\bar{N}}(\alpha) \mid N^+=\bar{N}-1 \right) &= p \cdot \mathbb{P}\left(T<b_{\bar{N}}(\alpha) \mid N^{+}=\bar{N}-1, N=\bar{N}\right) + (1-p) \\
&= p \cdot [\mathbb{P} \left( \alpha_i<\alpha \mid \alpha_i>\alpha^* \right)]^{\bar{N}-1} + (1-p) \\
&= p \left(\frac{\alpha-\alpha^*}{1-\alpha^*}\right)^{\bar{N}-1} + (1-p),
\end{aligned}
\]
which implies
\[
b_{\bar{N}}(\alpha) = T_{\bar{N}-1}\left(p\left(\frac{\alpha-\alpha^*}{1-\alpha^*}\right)^{\bar{N}-1}+(1-p)\right).
\]
Taking derivative, we get
\[
b_{\bar{N}}'(\alpha) = T_{\bar{N}-1}'\left(p\left(\frac{\alpha-\alpha^*}{1-\alpha^*}\right)^{\bar{N}-1}+(1-p)\right) \cdot p(\bar{N}-1)\left(\frac{\alpha-\alpha^*}{1-\alpha^*}\right)^{\bar{N}-2} \frac{1}{1-\alpha^*} .
\]
Let $\alpha=1$, we then have
\[
T_{\bar{N}-1}'(1) \cdot p(\bar{N}-1) \frac{1}{1-\alpha^*} = b_{\bar{N}}'(1) = T_{\bar{N}}^{\prime}(1) \cdot \bar{N} \frac{1}{1-\alpha^*},
\]
which implies
\[
p = \frac{\bar{N} T_{\bar{N}}'(1)}{(\bar{N}-1) T_{\bar{N}-1}'(1)}.
\]
Since RHS is observed, $p$ is identified. Note that
\[
p = \mathbb{P}(N=\bar{N} \mid N^{\text{act}}=\bar{N}-1) = \frac{\mathbb{P}(N=\bar{N},N^{\text{act}}=\bar{N}-1)}{\mathbb{P}(N^{\text{act}}=\bar{N}-1)}
\]
and
\[
\begin{aligned}
\frac{\mathbb{P}(N^+=\bar{N}-1)}{\mathbb{P}(N^+=\bar{N})} = \frac{\mathbb{P}(N^{\text{act}}=\bar{N}-1 \mid N^{\text{act}} \geq 1)}{\mathbb{P}(N^{\text{act}}=\bar{N} \mid N^{\text{act}} \geq 1)} = \frac{\mathbb{P}(N^{\text{act}}=\bar{N}-1)}{\mathbb{P}(N^{\text{act}}=\bar{N})},
\end{aligned}
\]
thus
\[
\begin{aligned}
p \cdot \frac{\mathbb{P}(N^+=\bar{N}-1)}{\mathbb{P}(N^+=\bar{N})} &= \frac{\mathbb{P}(N=\bar{N},N^{\text{act}}=\bar{N}-1)}{\mathbb{P}(N^{\text{act}}=\bar{N})} \\
&= \frac{\mathbb{P}(N=\bar{N})\mathbb{P}(N^{\text{act}}=\bar{N}-1 \mid N=\bar{N})}{\mathbb{P}(N=\bar{N})\mathbb{P}(N^{\text{act}}=\bar{N} \mid N=\bar{N})} \\
&= \frac{\bar{N}\alpha^*(1-\alpha^*)^{\bar{N}-1}}{(1-\alpha^*)^{\bar{N}}} \\
&= \frac{\bar{N}\alpha^*}{1-\alpha^*}
\end{aligned}
\]
Since 
\[
\frac{\sum_{l=1}^L \mathbb{1}(N_l^+=\bar{N}-1)}{\sum_{l=1}^L \mathbb{1}(N_l^+=\bar{N})} \xrightarrow{p} \frac{\mathbb{P}(N^+=\bar{N}-1)}{\mathbb{P}(N^+=\bar{N})}
\]
and $g(\alpha)=\frac{\alpha}{1-\alpha}$ is strictly increasing, $\alpha^*$ is identified. Therefore, $V(\alpha)$ for $\alpha \geq \alpha^*$ can be identified by
\[
\begin{aligned}
V(\alpha) &= b_{\bar{N}}(\alpha) + \frac{\alpha}{\bar{N}-1} b_{\bar{N}}'(\alpha) \\
&= T_{\bar{N}}\left(\left(\frac{\alpha-\alpha^*}{1-\alpha^*}\right)^{\bar{N}}\right) + \frac{\bar{N}}{\bar{N}-1}\left(\frac{\alpha-\alpha^*}{1-\alpha^*}\right)^{\bar{N}-1} \frac{\alpha}{1-\alpha^*} T_{\bar{N}}'\left(\left(\frac{\alpha-\alpha^*}{1-\alpha^*}\right)^{\bar{N}}\right).
\end{aligned}
\]

To prove non-identification for second-price auctions, it suffices to provide a counter-example. Denote the bidder's private value by $v_i$ and the reserve price by $R$. Consider the following two cases. In case 1, $N_l=2$, $v_i \sim \text{Unif}[0,1]$. $R=0.5$, which implies $\alpha^*=0.5$. Then we have
\[
\begin{aligned}
\mathbb{P}(N_l^+=1) &= \frac{2(1-\alpha^*)\alpha^*}{1-(\alpha^*)^2} = \frac{2}{3}, \\
\mathbb{P}(N_l^+=2) &= \frac{(1-\alpha^*)^2}{1-(\alpha^*)^2} = \frac{1}{3}. 
\end{aligned}
\]
Moreover, we have
\[
\mathbb{P}(T_l=0.5 \mid N_l^+=1) = 1,
\]
and
\[
\mathbb{P}(T_l<t \mid N_l^+=2) = \frac{(t-0.5)^2+2(t-0.5)(1-t)}{(1-0.5)^2} = -4t^2+8t-3 \ \ \forall t \in [0.5,1]. 
\]
In case 2, $N_l=1$ with probability $p_1=0.4$, $N_l=2$ with probability $p_2=0.6$, $v_i \sim \text{Unif}[\frac{1}{4},1]$. $R=0.5$, which implies $\alpha^*=\frac{1}{3}$. Then
\[
\begin{aligned}
\mathbb{P}(N_l^{\text{act}}=0) &= p_1\alpha^*+p_2(\alpha^*)^2 = \frac{1}{5}, \\
\mathbb{P}(N_l^{\text{act}}=1) &= p_1(1-\alpha^*)+p_2 \cdot 2(1-\alpha^*)\alpha^* = \frac{8}{15}, \\
\mathbb{P}(N_l^{\text{act}}=2) &= p_2(1-\alpha^*)^2 = \frac{4}{15}, \\
\end{aligned}
\]
which implies
\[
\begin{aligned}
\mathbb{P}(N_l^+=1) &= \frac{\mathbb{P}(N_l^{\text{act}}=1)}{\mathbb{P}(N_l^{\text{act}}=1)+\mathbb{P}(N_l^{\text{act}}=2)} = \frac{2}{3}, \\
\mathbb{P}(N_l^+=2) &= \frac{\mathbb{P}(N_l^{\text{act}}=2)}{\mathbb{P}(N_l^{\text{act}}=1)+\mathbb{P}(N_l^{\text{act}}=2)} = \frac{1}{3}.
\end{aligned}
\]
Besides, it can also be shown
\[
\mathbb{P}(T_l=0.5 \mid N_l^+=1) = 1,
\]
and
\[
\mathbb{P}(T_l<t \mid N_l^+=2) = \frac{(t-0.5)^2+2(t-0.5)(1-t)}{(1-0.5)^2} = -4t^2+8t-3 \ \ \forall t \in [0.5,1].
\]
Therefore, case 1 and case 2 generate the same joint distribution of $T_l$ and $N_l^+$, which implies they are not distinguishable from the data. In fact, any $\alpha^* \in (0,0.5)$ could rationalize this observed transaction price distribution with a proper $p \equiv \mathbb{P}(N_l=1)$. Note that  
\[
\begin{aligned}
&\mathbb{P}(N_l^{\text{act}}=1) = p(1-\alpha^*)+(1-p) \cdot 2\alpha^*(1-\alpha^*) \\
&\mathbb{P}(N_l^{\text{act}}=2) = (1-p) \cdot (1-\alpha^*)^2 \\
\end{aligned}
\]
Let $p=\frac{2-4\alpha^*}{3-4\alpha^*}$, then it can be shown 
\[
\mathbb{P}(N_l^{\text{act}}=1) = 2\mathbb{P}(N_l^{\text{act}}=2),
\]
which implies $\mathbb{P}(N_l^+=1)=\frac{2}{3}$ and $\mathbb{P}(N_l^+=2)=\frac{1}{3}$. Furthermore, suppose $v_i \sim \text{Unif}\left[\frac{0.5-\alpha^*}{1-\alpha^*},1\right]$, then it can be shown
\[
\begin{aligned}
&\mathbb{P}(T_l=0.5 \mid N_l^+=1) = 1, \\
&\mathbb{P}(T_l<t \mid N_l^+=2) = -4t^2+8t-3 \ \ \forall t \in [0.5,1],
\end{aligned}
\]
which is exactly the same as the previous case. \hfill $\blacksquare$ \\

\subsection*{Proof of Theorem 5}

Let
\[
P = \begin{pmatrix}
0 \\
\mathbb{P}(N=1) \\
\vdots \\
\mathbb{P}(N=\bar{N}) \\
\end{pmatrix}, \ \
P^{\text{obs}} = \frac{1}{L+L^{\text{invalid}}} \begin{pmatrix}
L^{\text{invalid}} \\
\sum_l \mathbb{1}(N_l^+=1) \\
\vdots \\
\sum_l \mathbb{1}(N_l^+=\bar{N}) \\
\end{pmatrix},
\]
and
\[
C(\alpha) = \begin{pmatrix}
1 & \alpha & \cdots & \alpha^{\bar{N}-1} & \alpha^{\bar{N}} \\
0 & C_1^1(1-\alpha) & \cdots & C_{\bar{N}-1}^1 (1-\alpha)\alpha^{\bar{N}-2} & C_{\bar{N}}^1 (1-\alpha)\alpha^{\bar{N}-1} \\
\vdots & \vdots & \ddots & \vdots &\vdots \\
0 & 0 & \cdots & C_{\bar{N}-1}^{\bar{N}-1} (1-\alpha)^{\bar{N}-1} & C_{\bar{N}}^{\bar{N}-1} (1-\alpha)^{\bar{N}-1}\alpha \\
0 & 0 & \cdots & 0 & C_{\bar{N}}^{\bar{N}} (1-\alpha)^{\bar{N}}\\
\end{pmatrix}.
\]
Then we have
\[
P^{\text{obs}} \xrightarrow{p} C(\alpha^*)P.
\]
There are $\bar{N}+1$ equations and $\bar{N}+1$ unknowns, thus $\alpha^*$ is at least set-identified and the identified set for $\alpha^*$ is 
\[
\left\{ \alpha \in [0,1] \mid \left( C(\alpha)^{-1}P^{\text{obs}} \right)_1=0, \left( C(\alpha)^{-1}P^{\text{obs}} \right)_j \in [0,1] \ \ \forall j=2,3,...,\bar{N}+1 \right\}.
\]
To show this set contains at most $\bar{N}$ elements, note the first row of $C(\alpha)^{-1}$ is 
\[
r(\alpha) = \begin{pmatrix}
1 & -\frac{\alpha}{1-\alpha} & \frac{\alpha^2}{(1-\alpha)^2} & \cdots & (-1)^{\bar{N}} \frac{\alpha^{\bar{N}}}{(1-\alpha)^{\bar{N}}} \\
\end{pmatrix}.
\]
Thus $(1-\alpha)^{\bar{N}}r(\alpha)P^{\text{obs}}$ is a polynomial of degree $\bar{N}$, which implies $r(\alpha)P^{\text{obs}}=0$ has at most $\bar{N}$ solutions in $(0,1)$. 

Let $T_{\bar{N}}(\cdot)$ denote the quantile function of the transaction price when $N^+=\bar{N}$, then $V(\alpha)$ for $\alpha \geq \alpha^*$ is also set-identified by
\[
V(\alpha) = T_{\bar{N}} \left( \phi_{\bar{N}} \left( \frac{\alpha-\alpha^*}{1-\alpha^*} \right) \right),
\]
where $\phi_{\bar{N}}(\alpha)=\bar{N}\alpha^{\bar{N}-1}-(\bar{N}-1)\alpha^{\bar{N}}$. By similar arguments to those in the proof of Theorem 1, it can be shown that all candidates in the identified set for $(\alpha,V(\alpha))$ could rationalize the distribution of $\left(T^{\text{obs}},N^+,L^{\text{invalid}}\right)$. Therefore, this identified set is sharp. \hfill $\blacksquare$

\subsection*{Proof of Theorem 6}

First consider first-price auctions. Denote the quantile function of the observed transaction price by $T(\cdot)$. Pick $\alpha^* \in (0,1)$ arbitrarily. For $\alpha \in [\alpha^*,1]$, define
\[
\begin{aligned}
&b(\alpha) = T \left( \frac{\alpha^N-(\alpha^*)^N}{1-(\alpha^*)^N} \right) \\
&V(\alpha) = b(\alpha) + \frac{\alpha}{N-1} b'(\alpha), \\
& c = [V(\alpha^*)-R] (\alpha^*)^{N-1} = \frac{b'(\alpha^*) (\alpha^*)^N}{N-1}.
\end{aligned}
\]
Using similar arguments in Lemma 1, it can be shown $T(\cdot)$ could be generated by $V(\alpha)$ with entry threshold $\alpha^*$ and entry cost $c$. Since the choice of $\alpha^*$ is arbitrary, the model is not identified.

To show $c$, $\alpha_N^*$, and $V(\alpha)$ for $\alpha \geq \alpha_N^*$ are identified in second-price auctions, note that
\[
\begin{aligned}
\frac{1}{L} \sum_{l=1}^L \mathbb{1}(T_l=R) &\xrightarrow{p} \mathbb{P}( N^{\text{act}}=1 \mid N^{\text{act}} \geq 1) \\
&= \frac{N(1-\alpha_N^*)(\alpha_N^*)^{N-1}}{1-(\alpha_N^*)^N}. \\
\end{aligned}
\]
Since RHS is a strictly monotone function of $\alpha_N^*$ and LHS is observed from the data, $\alpha_N^*$ is identified. Let $T_2(\cdot)$ denote the quantile function of $T_l \mid T_l>R$, then
$V(\alpha)$ for $\alpha \geq \alpha_N^*$ can be identified by 
\[
V(\alpha) = T_2 \left( \frac{\phi(\alpha)-\phi(\alpha_N^*)}{1-\phi(\alpha_N^*)} \right),
\]
and $c$ can be identified by 
\[
c = \left[ V(\alpha_N^*)-R \right] \cdot (\alpha_N^*)^{N-1} = \left[ T_2(0)-R \right] \cdot (\alpha_N^*)^{N-1},
\] 
where
\[
\phi(\alpha) = N\alpha^{N-1}-(N-1)\alpha^N.
\]
Therefore, $c$, $\alpha_N^*$, and $V(\alpha)$ for $\alpha \geq \alpha_N^*$ are all identified.  \hfill $\blacksquare$ \\

\subsection*{Proof of Theorem 7}

Using the arguments in Theorem 6,
\[
V(\alpha_N^*) = b(\alpha_N^*) + \frac{\alpha_N^*}{N-1}b'(\alpha_N^*) = R + \frac{N}{N-1} \frac{(\alpha_N^*)^N}{1-(\alpha_N^*)^N} T'(0).
\]
Thus the boundary condition can be rewritten as
\[
\left[ V(\alpha_N^*)-R \right] \cdot (\alpha_N^*)^{N-1} = \frac{N}{N-1} \frac{(\alpha_N^*)^N}{1-(\alpha_N^*)^N} T'(0) (\alpha_N^*)^{N-1} = c,
\]
which implies
\[
\frac{(\alpha_N^*)^{2N-1}}{1-(\alpha_N^*)^N} = \frac{N-1}{N} \frac{c}{T'(0)}.
\]
Since LHS is strictly monotone in $\alpha_N^*$ and RHS is observed from the data, $\alpha_N^*$ is point identified, and thus $V(\alpha)$ for $\alpha \geq \alpha_N^*$ is identified. \hfill $\blacksquare$ \\

\subsection*{Proof of Theorem 8}

Pick two different observed numbers of potential bidders $N_1$, $N_2$. Denote the entry thresholds at $N_1$ and $N_2$ by $\alpha_1$ and $\alpha_2$. Denote the quantile function of the observed transaction prices in auctions with $N_1$ and $N_2$ bidders by $T_1(\alpha)$ and $T_2(\alpha)$. Denote the corresponding equilibrium bidding function by $b_1(\alpha)$ and $b_2(\alpha)$. Then we have
\[
\begin{aligned}
&b_1(\alpha) = T_1 \left( \frac{\alpha^{N_1}-\alpha_1^{N_1}}{1-\alpha_1^{N_1}} \right) \ \ \forall \alpha \geq \alpha_1, \\
&b_2(\alpha) = T_2 \left( \frac{\alpha^{N_2}-\alpha_2^{N_2}}{1-\alpha_2^{N_2}} \right)  \ \ \forall \alpha \geq \alpha_2, \\
\end{aligned}
\]
which implies
\[
\begin{aligned}
V(\alpha) &= b_1(\alpha) + \frac{\alpha}{N_1-1}b_1'(\alpha) \\
&= T_1 \left( \frac{\alpha^{N_1}-\alpha_1^{N_1}}{1-\alpha_1^{N_1}} \right) + \frac{\alpha}{N_1-1} \frac{N_1\alpha^{N_1-1}}{1-\alpha_1^{N_1}} T_1' \left( \frac{\alpha^{N_1}-\alpha_1^{N_1}}{1-\alpha_1^{N_1}} \right),
\end{aligned}
\]
and
\[
\begin{aligned}
V(\alpha) &= b_2(\alpha) + \frac{\alpha}{N_2-1}b_2'(\alpha) \\
&= T_2 \left( \frac{\alpha^{N_2}-\alpha_2^{N_2}}{1-\alpha_2^{N_2}} \right) + \frac{\alpha}{N_2-1} \frac{N_2\alpha^{N_2-1}}{1-\alpha_2^{N_2}} T_2' \left( \frac{\alpha^{N_2}-\alpha_2^{N_2}}{1-\alpha_2^{N_2}} \right).
\end{aligned}
\]
Let $\alpha=1$, we have\
\[
T_1(1) + \frac{1}{N_1-1} \frac{N_1}{1-\alpha_1^{N_1}} T_1'(1) = V(1) = T_2(1) + \frac{1}{N_2-1} \frac{N_2}{1-\alpha_2^{N_2}} T_2'(1).
\]
Let $\alpha=\alpha_1$, we have
\[
V(\alpha_1) = T_1(0) + \frac{N_1}{N_1-1} \frac{\alpha_1^{N_1}}{1-\alpha_1^{N_1}} T_1'(0).
\]
Let
$\alpha=\alpha_2$, we have
\[
V(\alpha_2) = T_2(0) + \frac{N_2}{N_2-1} \frac{\alpha_2^{N_2}}{1-\alpha_2^{N_2}} T_2'(0).
\]
Since 
\[
c = [V(\alpha_1)-R] \cdot \alpha_1^{N_1-1} = [V(\alpha_2)-R] \cdot \alpha_2^{N_2-1},
\]
and $T_1(0)=T_2(0)=R$, we have
\[
\frac{N_1}{N_1-1} \frac{\alpha_1^{N_1}}{1-\alpha_1^{N_1}} T_1'(0)  \cdot \alpha_1^{N_1-1} = \frac{N_2}{N_2-1} \frac{\alpha_2^{N_2}}{1-\alpha_2^{N_2}} T_2'(0) \cdot \alpha_2^{N_2-1}.
\]
Combine them together, we have
\[
\left\{ 
\begin{aligned}
&T_1(1) + \frac{N_1}{N_1-1} \frac{1}{1-\alpha_1^{N_1}} T_1'(1) = T_2(1) + \frac{N_2}{N_2-1} \frac{1}{1-\alpha_2^{N_2}} T_2'(1), \\
&\frac{N_1}{N_1-1} \frac{\alpha_1^{2N_1-1}}{1-\alpha_1^{N_1}} T_1'(0) = \frac{N_2}{N_2-1} \frac{\alpha_2^{2N_2-1}}{1-\alpha_2^{N_2}} T_2'(0).
\end{aligned}
\right.
\]
Define
\[
x_j=\frac{1}{1-\alpha_j^{N_j}}, \qquad j=1,2.
\]
Since $\alpha_j\in(0,1)$, we have $x_j>1$. Also define
\[
A_j=\frac{N_j}{N_j-1}T_j'(1),
\qquad
B_j=\frac{N_j}{N_j-1}T_j'(0).
\]
By Lemma 1 and $c>0$, we have $A_j>0$ and $B_j>0$ for $j=1,2$. Then the first equation can be written as
\[
T_1(1)+A_1x_1=T_2(1)+A_2x_2,
\]
which implies
\[
x_2=a+bx_1,
\]
where
\[
a=\frac{T_1(1)-T_2(1)}{A_2},
\qquad
b=\frac{A_1}{A_2}>0.
\]
Note that $T_j(1)=b_j(1)$. Since $N_1>N_2$, we have $b_1(1)>b_2(1)$, and therefore $a>0$. Next, since
\[
\alpha_j^{N_j}=1-\frac{1}{x_j}=\frac{x_j-1}{x_j},
\]
we have
\[
\frac{\alpha_j^{2N_j-1}}{1-\alpha_j^{N_j}}
=
\frac{(x_j-1)^{\frac{2N_j-1}{N_j}}}
{x_j^{\frac{N_j-1}{N_j}}}.
\]
Thus the second equation becomes
\[
B_1
\frac{(x_1-1)^{\frac{2N_1-1}{N_1}}}
{x_1^{\frac{N_1-1}{N_1}}}
=
B_2
\frac{(x_2-1)^{\frac{2N_2-1}{N_2}}}
{x_2^{\frac{N_2-1}{N_2}}}.
\]
Raising both sides to the power $N_1N_2$ and rearranging gives
\[
\begin{aligned}
0=P(x_1)\equiv\;&
B_1^{N_1N_2}
(x_1-1)^{(2N_1-1)N_2}
x_2^{(N_2-1)N_1} \\
&-
B_2^{N_1N_2}
(x_2-1)^{(2N_2-1)N_1}
x_1^{(N_1-1)N_2}.
\end{aligned}
\]
Substituting $x_2=a+bx_1$, we obtain
\[
\begin{aligned}
P(x_1)=\;&
B_1^{N_1N_2}
(x_1-1)^{(2N_1-1)N_2}
(a+bx_1)^{(N_2-1)N_1} \\
&-
B_2^{N_1N_2}
(a+bx_1-1)^{(2N_2-1)N_1}
x_1^{(N_1-1)N_2}.
\end{aligned}
\]
Since all exponents are positive integers, $P(x_1)$ is a polynomial in $x_1$. Moreover,
\[
P(0)
=
B_1^{N_1N_2}
(-1)^{(2N_1-1)N_2}
a^{(N_2-1)N_1}
\neq 0,
\]
because $B_1>0$ and $a>0$. Therefore, $P$ is not the zero polynomial and has only finitely many roots.

Since $x_2=a+bx_1$, each root $x_1$ determines a unique $x_2$. Moreover,
\[
\alpha_j
=
\left(1-\frac{1}{x_j}\right)^{1/N_j},
\qquad j=1,2,
\]
so each pair $(x_1,x_2)$ determines a unique pair $(\alpha_1,\alpha_2)$. Hence the system has only finitely many solutions $(\alpha_1,\alpha_2)\in(0,1)^2$.

For each such pair, the common entry cost $c$ is uniquely determined by
\[
c=[V(\alpha_1)-R]\alpha_1^{N_1-1}
=[V(\alpha_2)-R]\alpha_2^{N_2-1},
\]
and $V(\alpha)$ for $\alpha>\min\{\alpha_1,\alpha_2\}$ is uniquely recovered from the observed transaction price distributions. Therefore, $c$ and $V(\alpha)$ for $\alpha$ above some threshold are set identified, with a finite identified set. \hfill $\blacksquare$

\subsection*{Proof of Theorem 9}

\textit{Proof.} First consider first-price auctions. Denote the equilibrium entry threshold and bidding strategy when the number of potential bidders equals to $N$ by $\alpha_N^*$ and $b_N(\alpha)$. From Lemma 1, $b_N(\alpha)$ is given by
\[
b_N(\alpha) = V(\alpha) - \frac{1}{\alpha^{N-1}} \left[ c + \int_{\alpha_N^*}^{\alpha} V'(t)t^{N-1} dt \right],
\] 
and thus
\[
b_N(1) = V(1) - \left[ c + \int_{\alpha_N^*}^{1} V'(t)t^{N-1} dt \right].
\]
Since $\alpha_N^*$ increases with $N$, $b_N(1)$ also increases with $N$. Note that $\bar{N}$ is identified by $\max\{N_l^+\}$ and $b_{\bar{N}}(1)$ is identified by $\max \left\{ T_l \mid N^+=\bar{N} \right\}$. Since $N^+=\bar{N}$ implies $N=\bar{N}$, for $t\in [b_{\bar{N}-1}(1),b_{\bar{N}}(1)]$, we have
\[
\mathbb{P}(T>t \mid N^+=\bar{N}) = 1 - \left[ \mathbb{P}(b_{\bar{N}}(\alpha_i)<t \mid \alpha_i>\alpha_{\bar{N}}^*) \right]^{\bar{N}},
\]
and
\[
\mathbb{P}(T>t \mid N^+=\bar{N}-1) = p \left\{ 1 - \left[ \mathbb{P}(b_{\bar{N}}(\alpha_i)<t \mid \alpha_i>\alpha_{\bar{N}}^*) \right]^{\bar{N}-1} \right\},
\]
where $p=\mathbb{P}(N=\bar{N} \mid N^+=\bar{N}-1)$. Since the LHS of these two equations are observed, $p$ can be identified by letting $t \uparrow b_{\bar{N}}(1)$. Note that
\[
p = \mathbb{P}(N=\bar{N} \mid N^{\text{act}}=\bar{N}-1) = \frac{\mathbb{P}(N^{\text{act}}=\bar{N}-1,N=\bar{N})}{\mathbb{P}(N^{\text{act}}=\bar{N}-1)},
\]
and
\[
\frac{\mathbb{P}(N^+=\bar{N}-1)}{\mathbb{P}(N^+=\bar{N})} = \frac{\mathbb{P}(N^{\text{act}}=\bar{N}-1 \mid N^{\text{act}} \geq 1)}{\mathbb{P}(N^{\text{act}}=\bar{N} \mid N^{\text{act}} \geq 1)} = \frac{\mathbb{P}(N^{\text{act}}=\bar{N}-1)}{\mathbb{P}(N^{\text{act}}=\bar{N})},
\]
thus
\[
\begin{aligned}
p \cdot \frac{\mathbb{P}(N^+=\bar{N}-1)}{\mathbb{P}(N^+=\bar{N})} &= \frac{\mathbb{P}(N^{\text{act}}=\bar{N}-1,N=\bar{N})}{\mathbb{P}(N^{\text{act}}=\bar{N})} \\
&= \frac{\mathbb{P}(N=\bar{N})\mathbb{P}(N^{\text{act}}=\bar{N}-1 \mid N=\bar{N})}{\mathbb{P}(N=\bar{N})\mathbb{P}(N^{\text{act}}=\bar{N} \mid N=\bar{N})} \\
&= \frac{\bar{N}\alpha_{\bar{N}}^*}{1-\alpha_{\bar{N}}^*}.
\end{aligned}
\]
Since LHS is observed, RHS is identified, which implies $\alpha_{\bar{N}}^*$ is identified. Therefore, $c$ and $V(\alpha)$ for $\alpha \geq \alpha_{\bar{N}}^*$ can be identified by similar arguments in Theorem 6.

Now consider second-price auctions. It suffices to provide a counter-example. Consider the following two cases: In case 1, $N_l=2$, $v_i \sim \text{Unif}[0,1]$, $c=0.36$, $R=0$, and thus $\alpha^*=0.6$. Then we have
\[
\begin{aligned}
\mathbb{P}(N_l^+=1) &= \frac{2(1-\alpha^*)\alpha^*}{1-(\alpha^*)^2} = \frac{3}{4}, \\
\mathbb{P}(N_l^+=2) &= \frac{(1-\alpha^*)^2}{1-(\alpha^*)^2} = \frac{1}{4}. 
\end{aligned}
\]
Moreover, since $R=0$, $V(\alpha^*)=0.6$, it can be shown
\[
\begin{aligned}
&\mathbb{P}(T_l=0 \mid N_l^+=1) = 1, \\
&\mathbb{P}(T_l<t \mid N_l^+=2) = 1-\left( \frac{1-t}{1-0.6} \right)^2 \ \ \forall t \in [0.6,1]. 
\end{aligned}
\]
In case 2, $N_l=1$ with probability $p_1=\frac{2}{9}$, $N_l=2$ with probability $p_2=\frac{7}{9}$, $v_i \sim \text{Unif}[0.2,1]$, $c=0.3$, $R=0$. Let $\alpha_1^*$, $\alpha_2^*$ denote the entry threshold when $N=1$ and $N=2$, respectively. Then it can be shown $\alpha_1^*=\frac{1}{8}$, $\alpha_2^*=0.5$. Therefore,
\[
\begin{aligned}
\mathbb{P}(N_l^{\text{act}}=0) &= p_1\alpha_1^*+p_2(\alpha_2^*)^2 = \frac{2}{9}, \\
\mathbb{P}(N_l^{\text{act}}=1) &= p_1(1-\alpha_1^*)+p_2 \cdot 2\alpha_2^*(1-\alpha_2^*) = \frac{7}{12}, \\
\mathbb{P}(N_l^{\text{act}}=2) &= p_2(1-\alpha_2^*)^2 = \frac{7}{36}, \\
\end{aligned}
\]
which implies
\[
\begin{aligned}
\mathbb{P}(N_l^+=1) &= \frac{\mathbb{P}(N_l^{\text{act}}=1)}{\mathbb{P}(N_l^{\text{act}}=1)+\mathbb{P}(N_l^{\text{act}}=2)} = \frac{3}{4}, \\
\mathbb{P}(N_l^+=2) &= \frac{\mathbb{P}(N_l^{\text{act}}=2)}{\mathbb{P}(N_l^{\text{act}}=1)+\mathbb{P}(N_l^{\text{act}}=2)} = \frac{1}{4}.
\end{aligned}
\]
Besides, since $R=0$, $V(\alpha^*)=0.6$, it can also be shown
\[
\begin{aligned}
&\mathbb{P}(T_l=0 \mid N_l^+=1) = 1, \\
&\mathbb{P}(T_l<t \mid N_l^+=2) = 1-\left( \frac{1-t}{1-0.6} \right)^2 \ \ \forall t \in [0.6,1]. 
\end{aligned}
\]
Therefore, case 1 and case 2 generate the same joint distribution of $T_l$ and $N_l^+$, which implies they are not distinguishable from the data. In fact, any $v_i \sim \text{Unif}[u,1]$ for $u \in \left(0,\frac{4-\sqrt{7}}{5}\right)$ could rationalize this observed transaction price distribution with a proper $(p_1,p_2,c)$, where $p_1=\mathbb{P}(N=1)$, $p_2=\mathbb{P}(N=2)$. To see why, let
\[
\begin{aligned}
&\alpha_2^* = \frac{0.6-u}{1-u}, \\
&\alpha_1^* = \frac{0.6\alpha_2^*-u}{1-u}, \\
&c = 0.6\alpha_2^*, \\
&p_1 = \frac{(1-\alpha_2^*)(3-5\alpha_2^*)}{(1-\alpha_1^*)+(1-\alpha_2^*)(3-5\alpha_2^*)}, \\
&p_2 = \frac{1-\alpha_1^*}{(1-\alpha_1^*)+(1-\alpha_2^*)(3-5\alpha_2^*)}, \\
\end{aligned}
\]
then it can be shown $\alpha_1^*,\alpha_2^*,p_1,p_2 \in (0,1)$ and
\[
\begin{aligned}
&\mathbb{P}(N_l^+=1) = \frac{\mathbb{P}(N_l^{\text{act}}=1)}{\mathbb{P}(N_l^{\text{act}}=1)+\mathbb{P}(N_l^{\text{act}}=2)} = \frac{3}{4}, \\
&\mathbb{P}(N_l^+=2) = \frac{\mathbb{P}(N_l^{\text{act}}=2)}{\mathbb{P}(N_l^{\text{act}}=1)+\mathbb{P}(N_l^{\text{act}}=2)} = \frac{1}{4}, \\
&\mathbb{P}(T_l=0 \mid N_l^+=1) = 1, \\
&\mathbb{P}(T_l<t \mid N_l^+=2) = 1-\left( \frac{1-t}{1-0.6} \right)^2 \ \ \forall t \in [0.6,1], \\
\end{aligned}
\]
which completes the proof. \hfill $\blacksquare$

\subsection*{Proof of Theorem 10}

First I show the distribution of $N$ is identified. Let $\alpha_N^*$ denote the entry threshold when the number of potential bidders is $N$. Note that for $N \geq 2$, $V(\alpha_N^*)$ can be identified by $\min\{T_l \mid N_l^+=N\}$. Thus $\alpha_N^*$ can be identified from
$\left[ V(\alpha_N^*)-R \right] \cdot\left(\alpha_N^*\right)^{N-1} = c$. Let
\[
P = \begin{pmatrix}
\mathbb{P}(N=2) \\
\mathbb{P}(N=3) \\
\vdots \\
\mathbb{P}(N=\bar{N}) \\
\end{pmatrix}, \ \
P^+ = \begin{pmatrix}
\mathbb{P}(N^+=2) \\
\mathbb{P}(N^+=3) \\
\vdots \\
\mathbb{P}(N^+=\bar{N}) \\
\end{pmatrix}, \ \
P^{\text{act}} = \begin{pmatrix}
\mathbb{P}(N^{\text{act}}=2) \\
\mathbb{P}(N^{\text{act}}=3) \\
\vdots \\
\mathbb{P}(N^{\text{act}}=\bar{N}) \\
\end{pmatrix}, 
\]
and
\[
C(\alpha^*) = \begin{pmatrix}
C_2^2 (1-\alpha_2^*)^2 & C_3^2 (1-\alpha_3^*)^2\alpha_3^* & \cdots & C_{\bar{N}}^2 (1-\alpha_{\bar{N}}^*)^2 (\alpha_{\bar{N}}^*)^{{\bar{N}}-2} \\
0 & C_3^3 (1-\alpha_3^*)^3 & \cdots & C_{\bar{N}}^3 (1-\alpha_{\bar{N}}^*)^3 (\alpha_{\bar{N}}^*)^{{\bar{N}}-3} \\
\vdots & \vdots & \ddots & \vdots \\
0 & 0 & \cdots & C_{\bar{N}}^{\bar{N}} (1-\alpha_{\bar{N}}^*)^{\bar{N}} \\
\end{pmatrix}.
\]
Then we have
\[
P^+ = \frac{1}{\mathbb{P}(N^{\text{act}} \geq 1)} P^{\text{act}} = \frac{1}{\mathbb{P}(N^{\text{act}} \geq 1)} C(\alpha^*)P
\]
Since $C(\alpha^*)$ is upper triangular with non-zero diagonal entries, it is invertible. Thus
\[
P = \mathbb{P}(N^{\text{act}} \geq 1) \left[C(\alpha^*)\right]^{-1} P^+.
\]
Let $e=(1,1,...,1)$, $\mathbb{P}(N^{\text{act}} \geq 1)$ can be identified by $eP=1$. Therefore, $P$ is identified. 

To show $V(\alpha)$ for $\alpha \geq \alpha_{\bar{N}}^*$ is identified, we simply need to focus on auctions where $N^+=\bar{N}$. Denote the quantile function of the observed transaction price when $N^+=\bar{N}$ by $T_{\bar{N}}(\cdot)$. By similar arguments in the proof of Theorem 5, we have
\[
V(\alpha) = T_{\bar{N}}\left( \bar{N}\left(\frac{\alpha-\alpha_{\bar{N}}^*}{1-\alpha_{\bar{N}}^*}\right)^{\bar{N}-1} - (\bar{N}-1) \left(\frac{\alpha-\alpha_{\bar{N}}^*}{1-\alpha_{\bar{N}}^*}\right)^{\bar{N}} \right),
\]
which completes the proof. \hfill $\blacksquare$

\bibliographystyle{apalike}
\bibliography{ref}

\end{document}